\documentclass[aps,pra,showpacs,twocolumn]{revtex4-1}
\usepackage{amsfonts}
\usepackage{amsmath}
\usepackage{amssymb}
\usepackage{graphicx}%
\usepackage{epsfig}
\providecommand{\U}[1]{\protect\rule{.1in}{.1in}}

\newcommand{\rmi}{\mathrm{i}}

\begin{document}
\preprint{ }
\title{Above threshold ionization by few-cycle spatially inhomogeneous fields}
\author{M. F. Ciappina$^{1}$}
\author{J. A. P\'erez-Hern\'andez$^{2}$}
\author{T. Shaaran$^{1}$}
\author{J. Biegert$^{1,3}$}
\author{R. Quidant$^{1,3}$}
\author{M. Lewenstein$^{1,3}$}
\affiliation{$^{1}$ICFO-Institut de Ci\`ences Fot\`oniques, Mediterranean Technology Park, 08860 Castelldefels (Barcelona), Spain}
\affiliation{$^{2}$Centro de L\'aseres Pulsados CLPU, E-37008 Salamanca, Spain}
\affiliation{$^{3}$ICREA-Instituci\'o Catalana de Recerca i Estudis Avan\c{c}ats, Lluis Companys 23, 08010 Barcelona, Spain}

\begin{abstract}
We present theoretical studies of above threshold ionization (ATI) produced by spatially inhomogeneous fields. This kind of field appears as a result of the illumination of plasmonic nanostructures and metal nanoparticles with a short laser pulse. We use the time-dependent Schr\"odinger equation (TDSE) in reduced dimensions to understand and characterize the ATI features in these fields. It is demonstrated that the inhomogeneity of the laser electric field plays an important role in the ATI process and it produces appreciable modifications to the energy-resolved photoelectron spectra. In fact, our numerical simulations reveal that high energy electrons can be generated. Specifically, using a linear approximation for the spatial dependence of the enhanced plasmonic field and with a near infrared laser with intensities in the mid- $10^{14}$ W/cm$^{2}$ range, we show it is possible to drive electrons with energies in the near-keV regime. Furthermore, we study how the carrier envelope phase influences the emission of ATI photoelectrons for few-cycle pulses. Our quantum mechanical calculations are supported by their classical counterparts.
\end{abstract}
\maketitle

\section{Introduction}

In the field of the interaction of laser fields with matter, above-threshold ionization (ATI) has been a particularly interesting subject in both experimental and theoretical physics. ATI, which was  experimentally observed more than 30 years ago~\cite{Agostini1979}, occurs when an atom or molecule absorbs more photons than the minimum number required to ionize it, with the leftover energy being converted to the kinetic energy of the released electron.

With recent advances in laser technology, it has become possible to generate few-cycle pulses, which find a wide range of applications in science, such as controlling chemical reactions and molecular motion~\cite{schnurer2000,vdHoff2009}, and generating high-order harmonics and even the creation of isolated extreme ultraviolet (XUV) pulses~\cite{ferrari2010,schultze2007}.  These allow even more control on an attosecond temporal scale.

The electric field in a few-cycle pulse can be characterized by its duration and by the so-called carrier-envelope phase (CEP). In comparison to a multicycle pulse, the electric field of few-cycle pulses is greatly affected by the CEP~\cite{Wittmann2009,kling2008}. The influence of CEP has been experimentally observed in high-harmonic generation (HHG)~\cite{nisoli2003}, the emission direction of electrons from atoms~\cite{paulus2001} and in the yield of nonsequential double ionization~\cite{liu2004}. In order to have a better control of the system on an attosecond temporal scale it is, therefore, important to find reliable schemes to measure the absolute phase of few-cycle pulses.

Recently, the investigation of ATI generated by few-cycle driving laser pulses has attracted so much interest due to the sensitivity of the energy and angle-resolved photoelectron spectra to the absolute value of the CEP~\cite{paulus_cleo,sayler}. Consequently, this feature renders the ATI phenomenon a very valuable tool for laser pulse characterization. In order to characterize the CEP of a few-cycle laser pulse, the so-called backward-forward asymmetry of the ATI spectrum is measured and from the information collected the absolute CEP can be obtained~\cite{paulus}. Furthermore, nothing but the high energy region of the photoelectron spectra appears to be sensitive to the absolute CEP and consequently electrons with kinetic energy are needed in order to characterize it~\cite{milosevic_rev,paulus2003}.

New experiments have demonstrated that the harmonic cutoff and electron spectra of ATI could be extended further by using plasmon field enhancement~\cite{kim,kling}. This field appears when a metal nanostructure or nanoparticle is illuminated by a short laser pulse and it is not spatially homogeneous, due to the strong confinement of the plasmonics spots and the distortion of the electric field by the surface plasmons induced in the nanosystem.  One should note, however, that a recent controversy about the outcome of the experiments of Ref.~\cite{kim} has arisen~\cite{sivis,Kimreply,corkum_priv}. Consequently, alternative systems to the metal bow-tie shaped nanostructures have appeared~\cite{Kimnew}. A related process employing solid state targets instead of atoms and molecules in gas phase is the so called Above Threshold Photoemission (ATP). This laser driven phenomenon has received special attention recently due to its novelty and considering new physics could be involved. In ATP electrons are emitted from metallic surfaces or metal nanotips and they present distinct characteristics, namely higher energies, far beyond the usual cutoff for noble gases and consequently the possibility to reach similar electron energies with smaller laser intensities (see e.g.~\cite{peterprl2006,peterprl2010,peternature,peterjpbreview,jensdombi,ropers}). Furthermore, the photoelectrons emitted from these nanosources are sensitive to the CEP and consequently it plays an important role in the angle and energy resolved photoelectron spectra~\cite{apolonski,dombi,kling,peternature}. 

Despite new developments, all numerical and semiclassical approaches to model the ATI phenomenon are based on the assumption that the external field is spatially homogeneous in the region where the electron dynamics take place~\cite{keitel,krausz}. For an inhomogeneous field, however, important changes will occur to the features of strong field phenomena~\cite{kim,kling} since the laser-driven electric field, and consequently the force applied to the electron, will also depend on position. Up to now, there have been very few studies to investigate the strong field phenomena in such kind of fields~\cite{husakou,ciappi2012,yavuz,ciappi_prl}.  

From a theoretical viewpoint, the ATI process can be tackled using different approaches (for a summary see e.g.~\cite{milosevic_rev,schafer1993,telnov2009,bauer2006,Blaga2009,Quan2009} and references therein). In this article, we concentrate our effort in extending one of the most and widely used approaches: the numerical solution of time-dependent Schr\"odinger Equation (TDSE) in reduced dimensions. 
We have developed our numerical tool in such a way to allow the treatment of a very general set of nonhomogeneous fields. Furthermore, based on our model, we examine the influence of the CEP on photoelectron spectra of ATI. The kinetic energy for the rescattered electron is classically calculated and compared to our quantum mechanical approach. 

This article is organized as follows. In Sec.~II, we present our theoretical approach to model ATI produced by nonhomogeneous fields. Subsequently, in Sec.~III, we employ this method to compute the ATI energy-resolved photoelectron spectra using few-cycle laser pulses for both homogeneous and inhomogeneous fields. In addition, we perform classical simulations to support our quantum mechanical method. Finally, in Sec.~IV, we conclude with a short summary and outlook.

\section{Theoretical approach}

In order to calculate the energy resolved photoelectron spectra, we use the one-dimensional time-dependent Schr\"odinger
equation (1D-TDSE)
\begin{eqnarray}
\label{tdse}
\rmi \frac{\partial \Psi(x,t)}{\partial t}&=&\mathcal{H}(t)\Psi(x,t) \\
&=&\left[-\frac{1}{2}\frac{\partial^{2}}{\partial x^{2}}+V_{atom}(x)+V_{laser}(x,t)\right]\Psi(x,t) \nonumber
\end{eqnarray}
where $V_{laser}(x,t)$ represents the laser-atom interaction. For the atomic
potential, we use the quasi-Coulomb or soft core potential
\begin{eqnarray}  \label{atom}
V_{atom}(x)&=&-\frac{1}{\sqrt{x^2+a^2}}
\end{eqnarray}
which was introduced in~\cite{eberly} and has been widely used in the study
of laser-matter processes in atoms. The parameter $a$ in Eq. (\ref{atom})
allows us to match the ionization potential of the atom under consideration.
We consider the field to be linearly polarized along the $x$-axis and modify
the interaction term $V_{laser}(x,t)$ in order to treat spatially
nonhomogeneous fields, although maintaining the dipole character.
Consequently we write
\begin{eqnarray}  \label{vlaser}
V_{laser}(x,t)&=&-E(x,t)\,x
\end{eqnarray}
where $E(x,t)$ is the laser electric field defined as
\begin{equation}  \label{electric}
E(x,t)=E_0\,f(t)\, (1+\varepsilon h(x))\,\sin(\omega t+\phi).
\end{equation}
In Eq. (\ref{electric}), $E_0$, $\omega$ and $\phi$ are the peak
amplitude, the frequency of the laser pulse and the CEP, respectively. We refer to sin(cos)-like laser pulses where $\phi=0$ ($\phi=\pi/2$). The pulse envelope is given by $f(t)$ and $\varepsilon$
is a small parameter that characterizes the inhomogeneity strength. The function $h(x)$
represents the functional form of the nonhomogeneous field and, in
principle, could take any form and be supported by the numerical algorithm~\cite{ciappi_prl}. In this work, however, we concentrate our efforts on the
simplest form for $h(x)$, i.e. the linear term: $h(x)=x$. This choice is motivated by previous investigations in high-order harmonic generation~\cite{husakou,ciappi2012,yavuz,ciappi_prl,tahirsfa}.\footnote{The actual spatial dependence of the enhanced near-field in the surrounding of a metal nanostructure can be obtained by solving the Maxwell equations incorporating both the geometry and material properties of the nanosystem under study and the input laser pulse characteristics (see e.g.~\cite{ciappi_prl}). The electric field retrieved numerically is then approximated using a power series  $h(x) =\sum_{i=1}^{N}b_{i}x^{i}$, where the coefficients $b_i$ are obtained by fitting the real electric field that results from a finite element simulation. Furthermore, in the region relevant for the strong field physics and electron dynamics and in the range of the parameters we are considering, the electric field can be indeed approximated by its linear dependence.}

In the linear model we are using in this work, the units of $\varepsilon$ are inverse length (see also~\cite{husakou,yavuz,ciappi2012}). We have written $V_{laser}$ in Eq. (\ref{vlaser}) in such a way to emphasize the fact we are working within the dipole approximation and any deviation of it is considered small, i.e. higher electric multipole terms and magnetic effects are neglected~\cite{reiss}. To model short
laser pulses, we use a sin-squared envelope $f(t)$ of the form
\begin{equation}
f(t)=\sin^{2}\left(\frac{\omega t}{2 n_p}\right)
\end{equation}
where $n_p$ is the total number of optical cycles. The total duration of the
laser pulse will then be $T_p=n_p \tau$ where $\tau=2\pi/\omega$ is the laser
period.

We assume the target atom is in the ground state ($1s$) before we turn on the
laser ($t=-\infty$). This state can be found by solving an eigenvector and eigenvalue
problem once the spatial coordinate $x$ has been discretized. We chose $a^2=1.412$ to match the atomic ionization potential of our target, which is
an hydrogen atom ($I_p=-0.5$ a.u.).  Eq.(\ref{tdse}) is solved numerically by
using the Crank-Nicolson scheme with an adequate spatial grid~\cite{keitel}. We employ boundary reflections mask functions ~\cite{mask} in order to
avoid spurious contributions.

For calculating the energy-resolved photoelectron spectra $P(E)$ we use the
window function technique developed by Schafer~\cite{schaferwop1,schaferwop}. 
This tool has been widely used, both to calculate angle-resolved and
energy-resolved photoelectron spectra~\cite{schaferwop2} and it represents a
step forward with respect to the usual projection methods.

\section{Results}

In this section, we will determine the energy-resolved photoelectron spectra
$P(E)$ using Eq.~(\ref{tdse}), in order to investigate the role of the inhomogeneities
of the field. Furthermore, we demonstrate how the CEP $\phi$ will effect the
the energy-resolved photoelectron spectra of ATI.  We employ a four-cycle (total duration 10 fs) sin-squared laser pulse with an intensity 
$I=3\times10^{14}$ W/cm$^{2}$ and wavelength $\lambda=800$ nm.

We chose three different values for the parameter that characterizes the
inhomogeneity strength, namely $\varepsilon =0$ (homogeneous case), $0.003$ and $0.005$.  Figures 1 and 2
show the cases with $\phi =0$ (a sin-like laser pulse) and $\phi =\pi/2$
(a cos-like laser pulse), respectively. Panels (a) of both Figures represent
the homogeneous case, i.e. $\varepsilon =0$, and panels (b) and (c) show the
nonhomogeneous case with $\varepsilon =0.003$ and $\varepsilon =0.005$,
respectively.

\begin{figure}[htb]
\centering
\includegraphics[width=0.42\textwidth]{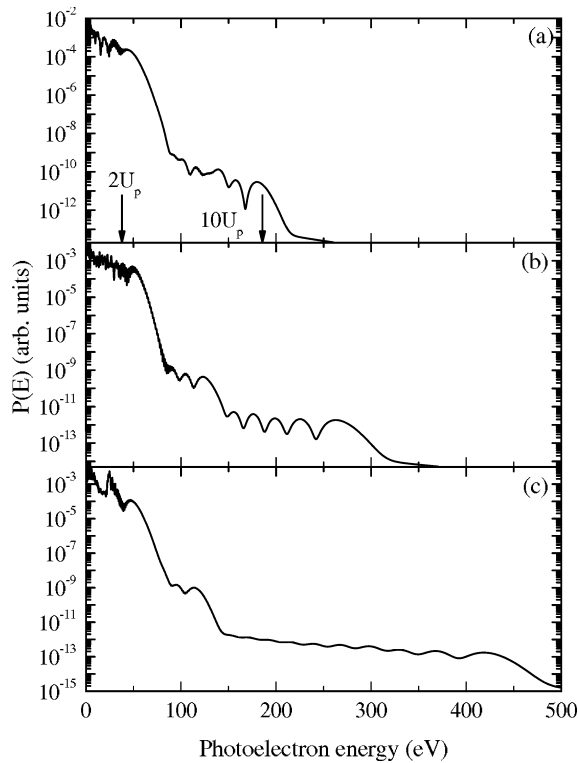}
\caption{Energy-resolved photoelectron spectra $P(E)$ calculated using the 1D-TDSE for a model atom with $I_{p}=-0.5$. The laser parameters are $I=3\times10^{14}$ W/cm$^{2}$ and $\lambda=800$ nm. We have used a sin-squared shaped pulse with a total duration of 4 cycles (10 fs) and $\phi=0$ (a sin-like pulse). The arrows indicate the $2 U_p$ and $10 U_p$ cutoffs predicted by the classical model~\cite{milosevic_rev}. Panel (a) $\varepsilon=0$ (homogeneous case), (b) $\varepsilon=0.003$ and (c) $\varepsilon=0.005$.
}
\label{fig:figure1}
\end{figure} 

For the homogeneous case, the spectra exhibits the usual distinct behavior,
namely the $2U_{p}$ cutoff ($\approx 36$ eV for our case) and the $10U_{p}$
cutoff ($\approx 180$ eV), where $U_{p}=E_{0}^{2}/4\omega^{2}$ is the
ponderomotive potential.  The former cutoff corresponds to those electrons
that, once ionized,  never return to the atomic core, while the latter one
corresponds to the electrons that, once ionized, return to the core and
elastically rescatter. It is well established using classical arguments that the maximum kinetic energies of the \textit{direct} and the \textit{rescattered} electrons are $E_{max}^{d}=2U_{p}$ and  $E_{max}^{r}=10U_{p}$,
respectively. In a quantum mechanical approach, however, it is possible to find electrons with energies beyond the 10$U_p$, although their yield of them drops several orders of magnitude~\cite{milosevic_rev}. Experimentally, both mechanisms contribute to the
energy-resolved photoelectron spectra and consequently the theoretical approach to tackle the problem
should to include them. In that sense the TDSE, which can be considered as an exact approach to the problem, is able to predict the $P(E)$ in the whole range of electron energies. In addition, the most energetic electrons, i.e. those with $E_{k}\gg 2U_{p}$, are used to
characterize the CEP of few-cycle pulses. As a result, a correct description
of the rescattering mechanism is needed.

\begin{figure}[htb]
\centering
\includegraphics[width=0.42\textwidth]{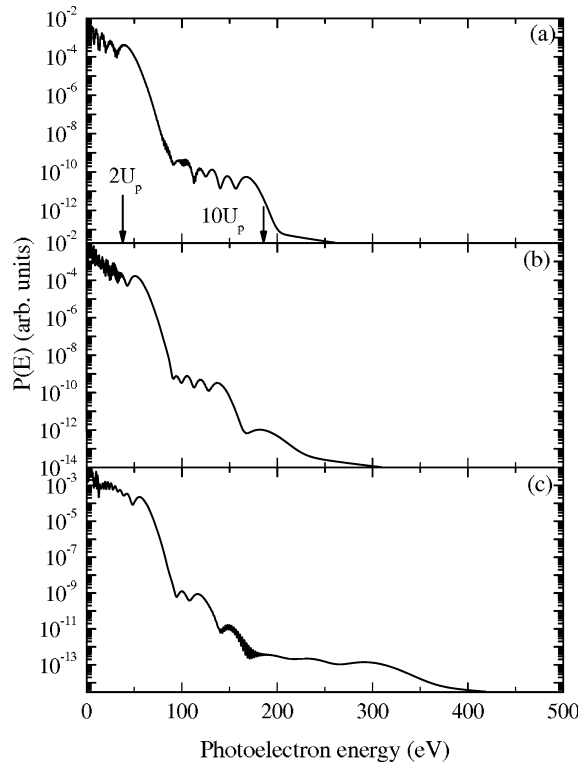}
\caption{Idem Fig. 1 but $\phi=\pi/2$ (a cos-like pulse).}
\label{fig:figure2}
\end{figure} 

For the inhomogeneous case, the cutoff positions of the \textit{direct} and the \textit{rescattered} electrons are extended towards larger energies.
For the \textit{rescattered} electrons, this extension  is very prominent. In
fact,  for $\varepsilon =0.003$ and  $\varepsilon =0.005,$ it reaches  $\approx 260$ eV and  $\approx 420$ eV (panels b and c of Fig. 1, respectively). 
Furthermore, it appears that the high energy region of $P(E)$, for instance,  the region between $200-400$ eV for $\varepsilon =0.005$
(see panels (c) of Figs. 1 and 2), is strongly sensitive to the CEP. This feature
indicates that the high energy region of the photoelectron spectra could resemble a new and better CEP characterization tool. It should be, however,
complemented by other well known and established CEP characterization tools, as, for instance, the forward-backward asymmetry (see~\cite{milosevic_rev}). 
Furthermore, the utilization of nonhomogeneous fields would open the avenue for the
production of high energy electrons, reaching the keV regime, if a reliable control of the spatial and temporal shape of the laser electric field is attained.

\begin{figure*}[htb]
\centering
\includegraphics[width=0.8\textwidth]{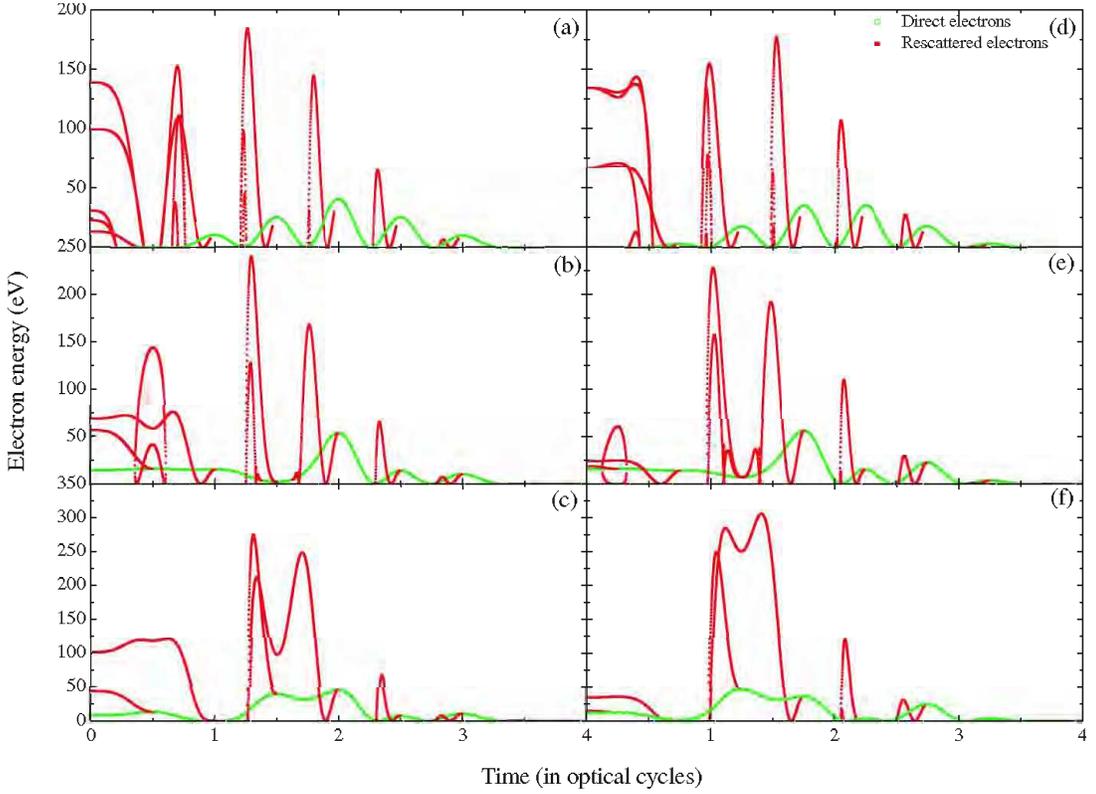}
\caption{(Color online). Numerical solutions of the Newton equation (Eq. (\ref{newton})) plotted in terms of the direct and rescattered electron kinetic energy. The laser parameters are the same as in Fig. 1.  Panels (a), (b) and (c) correspond to the case of sin-like pulses ($\phi=0$) and for $\varepsilon=0$ (homogeneous case), $\varepsilon=0.003$ and $\varepsilon=0.005$, respectively. Panels (d), (e) and (f) correspond to the case of cos-like pulses ($\phi=\pi/2$) and for $\varepsilon=0$ (homogeneous case), $\varepsilon=0.003$ and $\varepsilon=0.005$, respectively.}
\label{fig:figure3}
\end{figure*} 

We now concentrate our efforts in order to explain the extension of the
energy-resolved photoelectron spectra using classical arguments. From the
simple-man's model~\cite{corkum} we can describe the physical origin of the
ATI process as follows: an atomic electron at a position $x=0$, is released
or \textit{born} at a given time, that we call \textit{ionization} time $t_{i}$, 
with zero velocity, i.e. $\dot{x}(t_{i})=0$. This electron now moves
only under the influence of the oscillating laser electric field (the
residual Coulomb interaction is neglected in this model) and will reach
the detector either directly or through the rescattering process.  By using the
classical equation of motion, it is possible to calculate the maximum
energy of the electron for both direct and rescattered processes. The Newton
equation of motion for the electron in the laser field can be written as  (\ref{vlaser}):
\begin{eqnarray}
\ddot{x}(t) &=&-\nabla _{x}V_{laser}(x,t)  \notag  \label{newton} \\
&=&E(x,t)+\left[ \nabla _{x}E(x,t)\right] x  \notag \\
&=&E(t)(1+2\varepsilon x(t)),
\end{eqnarray}
where we have collected the time dependent part of the electric field in $E(t)$, i.e.  $E(t)=E_{0}f(t)\sin (\omega t+\phi )$ and we have
specialized to the case $h(x)=x$. In the limit where $\varepsilon =0$ in Eq.~(\ref{newton}), we recover the homogeneous case. For the
direct ionization, the kinetic energy of an electron released or born at
time $t_{i}$ is
\begin{equation}
\label{direct}
E_{d}=\frac{\left[ \dot{x}(t_{i})-\dot{x}(t_{f})\right] ^{2}}{2},
\end{equation}
where $t_{f}$ is the end time of the laser pulse. For the rescattered
ionization, in which the electron returns to the core at a time $t_{r}$ and
reverses its direction, the kinetic energy  of the electron yields
\begin{equation}
\label{rescattered}
E_{r}=\frac{\left[ \dot{x}(t_{i})+\dot{x}(t_{f})-2\dot{x}(t_{r})\right] ^{2}}{2}.
\end{equation}

For homogeneous fields, Eqs.~(\ref{direct}) and (\ref{rescattered}) become as $E_{d}=\frac{\left[ A(t_{i})-A(t_{f})\right] ^{2}}{2}$ and 
$E_{r}=\frac{\left[ A(t_{i})+A(t_{f})-2A(t_{r})\right] ^{2}}{2}$, with $A(t)$ being the laser vector potential $A(t)=-\int^{t} E(t')dt'$. For the case with $\varepsilon=0$, it
can be shown that the maximum value for $E_{d}$ is $2U_{p}$ while for $E_{r}$ it is $10U_{p}$~\cite{milosevic_rev}. These two values appear as cutoffs in the energy resolved photoelectron spectrum as can be
observed in panels (a) of Figs. 1 and 2 (see the respective arrows).

In Fig.~3, we present the numerical solutions of  Eq. (\ref{newton}), which is plotted in terms of the kinetic energy of the direct and rescattered
electrons. We employ the same laser parameters as in Figs. 1 and 2. Panels
(a), (b) and (c) correspond to the case of $\phi=0$ (sin-like pulses) and
for $\varepsilon=0$ (homogeneous case), $\varepsilon=0.003$ and $\varepsilon=0.005$, 
respectively. Meanwhile, panels (d), (e) and (f)
correspond to the case of $\phi=\pi/2$ (cos-like pulses) and for $\varepsilon=0$ (homogeneous case), 
$\varepsilon=0.003$ and $\varepsilon=0.005$, respectively.
From the panels (b), (c), (e) and (f) we can observe the strong
modifications that the nonhomogeneous character of the laser electric field
produces in the electron kinetic energy. These are related to the changes
in the electron trajectories (for details see e.g.~\cite{yavuz,ciappi2012,ciappi_prl}). In short, the electron trajectories are
modified in such a way that now the electron ionizes at an earlier time and
recombines later, and in this way it spends more time in the continuum
acquiring energy from the laser electric field. Consequently, higher values
of the kinetic energy are attained. A similar behavior with the photoelectrons was observed recently in ATP using metal nanotips. According to the model presented in Ref.~\cite{ropers} the localized fields modify the electron motion in such a way to allow sub-cycle dynamics. In our studies, however, we consider both direct and rescattered electrons (in Ref.~\cite{ropers} only direct electrons are modeled) and the characterization of the dynamics of the photoelectrons is more complex. Nevertheless, the higher kinetic energy of the rescattered electrons is a clear consequence of the strong modifications of the laser electric field in the region where the electron dynamics takes place, as in the above mentioned case of ATP.  

\section{Conclusions and Outlook}

We have extended previous studies of high-order harmonic generation produced
by nonhomogeneous fields to above threshold ionization (ATI). An example is the field generated in a vicinity of a metal nanostructure or nanoparticle when it is irradiated by a short laser pulse. We have modified the time dependent Schr\"odinger equation to model the ATI phenomenon driven
by nonhomogeneous fields. We predict an extension in the cutoff position and an increase of the yield of the energy-resolved photoelectron spectra in
certain regions. These features are reasonable well reproduced by classical simulations. Our predictions would pave the way to the production of high
energy photoelectrons, reaching the keV regime, using plasmon enhanced fields. Application of our model to a broader range of laser parameters,
including an exhaustive study of CEP effects, and a systematic survey over different atomic species using a full dimensional scheme will be subject of further investigations.

\section*{Acknowledgments}

We acknowledge the financial support of the MINCIN projects (FIS2008-00784 TOQATA and Consolider Ingenio 2010 QOIT) (M. F. C. and M.L.); ERC Advanced Grant QUAGATUA, Alexander von Humboldt Foundation and Hamburg Theory Prize (M. L.); Spanish MINECO (FIS2009-09522) (J. A. P-H); Spanish Ministry of Education and Science through its Consolider Program Science (SAUUL CSD 2007-00013), Plan Nacional (FIS2008-06368-C02-01),  LASERLAB-EUROPE (grant agreement n° 228334, EC's Seventh Framework Programme) (J. B.); this research has been partially supported by Fundaci\'o Privada Cellex. We thank Dane Austin for valuable comments and suggestions.


%
%
%

\end{document}